\documentstyle[art10,hip-100a,epsf]{article}
\def\f{\frac}
\def\v{\vec}
\begin{document}
\cimo \setcounter{page}{1} \thispagestyle{empty} \hskip -15pt  \noindent
{\Large\bf
Strangeness Production via Parton Cascade
}\\[3mm] 
\def\rightmark{Strangeness Production via Parton
Cascade}\def\leftmark{B. Zhang, et al.} 
\hspace*{6.327mm}\begin{minipage}[t]{12.0213cm}{\large\lineskip .75em
Bin Zhang$^1$, Miklos Gyulassy$^1$ and Yang Pang$^1$
}\\[2.812mm] 
\hspace*{-8pt}$^1$ Physics Department, Columbia University,\\
New York, NY 10027, USA\\[0.2ex]
\\[4.218mm]{\it
Received nn Month Year (to be given by the editors)
}\\[5.624mm]\noindent
{\bf Abstract.} We study  pre-equilibrium strangeness production at RHIC
energies in a new parton cascade. Starting with
the turbulent glue HIJING initial conditions we
investigate the interplay between
mini-jet and  soft beam jet gluons for strangeness
production prior to hadronization, and show the importance of soft beam jet
gluons in the strangeness production. 

\end{minipage}
 
\section{Introduction}

In ultrarelativistic heavy ion collisions, qualitatively new phenomena are
expected, among which the formation of quark gluon plasma (QGP). One of the
challenges for the nuclear physics community is to find signals that can
identify the formation of QGP. Based on pQCD and thermodynamic calculations,
strangeness enhancement was proposed long ago as 
one of the signals of QGP formation\cite{bib1}
due to the relatively fast strangeness production process $g+g\rightarrow
s+\bar{s}$. 

Strangeness production from more realistic initial conditions
(HIJING\cite{bib2} minijet initial conditions) has been investigated\cite{bib3}
 by following a set of rate equations in a 1-dimensional expanding system and
 it was out that the system does not reach chemical equilibrium.

Cascade Models can be used to investigate equilibration without the assumption
of local thermal equilibrium. Results from Klaus Geiger's Parton Cascade Code
(PCM)\cite{bib4} show that we can not get chemical equilibrium at RHIC
energies.

We investigate the strangeness production using a parton cascade code which we
developed
recently. In particular, we study the interplay between minijet gluons and soft
beam jet gluons and reveal the importance of soft beam jet gluons in the
strangeness production.

We'll first review the argument leading to fast equilibration,
followed by a simple discussion of our cascade approach. 

%
\vspace{0.7cm}
\hbox to\columnwidth{\hfil 0231-4428/95/ \$ 5.00}
\hbox to\columnwidth{\hfil
\copyright 1996 Akad\'emiai Kiad\'o, Budapest}
\newpage
\noindent
Then we discuss the differences between our cascade code and Geiger's PCM, 
and estimate
the relevant screening scales. Next
we present the result of our cascade
simulations which shows that chemical equilibrium can not be achieved and the
soft beam
jet gluons are important in the strangeness production.

\section{Parton cascade approach to strangeness production}

\subsection{Strangeness equilibration rate}
It's not difficult to get an estimate for the strangeness equilibration
rate. To solve the rate equation,
\[\f{dn_s}{dt}\approx A(1-(\f{n_s(t)}{n_s(\infty)})^2),\]
we need the transition rate per unit volume,
\[A=\f{dN}{dtd^3x}\]\[
=\f{1}{2}\int_{4m^2}^\infty\;ds
\;\sqrt{s(s-4m^2)}\,\delta(s-(k_1+k_2)^2)
\]\[\int\f{d^3k_1}{(2\pi)^3E_1}\int\f{d^3k_2}{(2\pi)^3E_2}
\f{1}{2}(2\times 8)^2 e^{-\beta (E_1+E_2)}\sigma_{gg\rightarrow s\bar{s}}(s),\]
in which $e^{-\beta E_1}$ and $e^{-\beta E_2}$ are classical phase space
distributions, and
\[\sigma_{gg\rightarrow s\bar{s}}(s)=\f{\pi \alpha^2}{3s}
\left[-(7+\f{31m^2}{s})\f{1}{4}\chi+(1+\f{4m^2}{s}+\f{m^4}{s^2})log\f{1+\chi}{1-\chi}\right]\]
where $\chi=\sqrt{1-\f{4m^2}{s}}$, and $\f{\sqrt{s(s-4m^2)}}{2E_1E_2}$ is the
relative velocity.

The relaxation time is given by:
\[\tau=n_s(\infty)/2A\]
and is shown in Fig.1 for $\alpha=0.6$. we see that the equilibration time is
around several 
fermis.

Since at central rapidity the freezeout time is also around several fermis, we
expect from the above simple estimate that chemical equilibrium may be achieved
during the life time of the plasma. Comparing to the relatively slower hadronic
strangeness production processes, strangeness signal will be enhanced by
the formation of quark gluon plasma.

\newpage
\vspace*{6cm}
\includegraphics{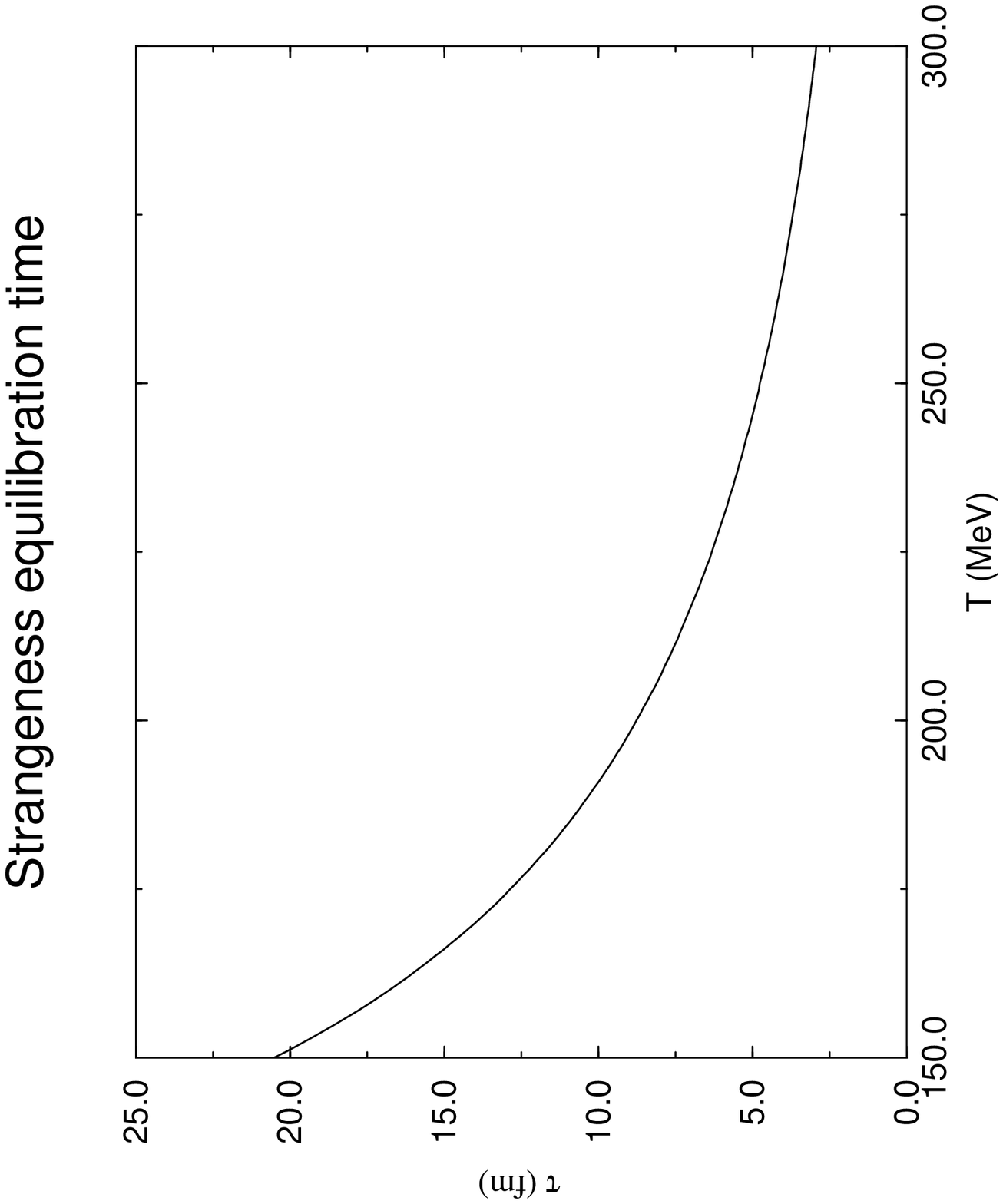}
\begin{center}
\hspace{0.5cm}\bf Fig.1.
\end{center}
\vskip 4truemm

\subsection{Parton Cascade Approach}

To study strangeness production at RHIC energies, without the assumption of
local thermal equilibrium necessary for the 
hydrodynamic approach, we develop our
own parton cascade code. This new cascade code is different from the
one Klaus Geiger developed in several important aspects.

Our initial conditions are taken from
HIJING output. 
We use inside-outside cascade picture in which particle production is a result
of the collision of two nuclei.
We do not have wee partons interacting before the collision of
two Lorentz contracted nuclei, because they are part of a coherent field before
the collision of the valance quarks. 

In our approach, we have on-shell particles instead of virtual particles so
that real parton scattering cross sections can be used. The
singularities of cross sections are regulated by a medium generated screening
mass; this differs from Geiger's approach in which a lower momentum cutoff is
used. 

For the soft beam jet gluons, we use 1000 soft gluons per unit rapidity
bin. They have a temperature parameter around 250 $MeV$ to account
for the beam energy loss at RHIC.

For parton parton collisions, we use two particle center of mass collision
criteria, i.e., two particle collide when their closest approach distance is
smaller than $\sqrt{\f{\sigma}{\pi}}$
($\sigma$ is the scattering cross section). The collision
point is chosen to be the midpoint of two particles in the two-body center of
mass 
frame at their closest approach point.

\newpage
We take leading divergent cross sections regulated by a screening mass except
for
$g+g\rightarrow s+\bar{s}$ which is regulated by the strangeness mass.

\subsection{Color screening in heavy ion collisions}

In nucleus-nucleus collisions, medium effects manifest as color screening. The
color screening mass is important in regulating the forward gluon-gluon
scattering cross section. Since the total elastic cross section depends on the
screening mass, it is an important factor in determining the strangeness
production branching ratio of the gluon-gluon scattering.

The screening mass is related to phase space distribution through\cite{bib5}:
\[m^2=-\f{3\alpha_s}{\pi^2}\lim_{|\v{q}|\rightarrow 0}
\int d^3k\f{|\v{k}|}{\v{q}\cdot\v{k}}\v{q}\cdot\v{\nabla}_{\v{k}}f(\v{k}),\]
in which $f(\vec{k})$ can be parametrized as:
\[f(\v{k})=\f{2(2\pi)^2}{g_GV}\f{1}{|\v{k}|}g(k_T,y),\]
where:
\[g(k_T,y)=\f{1}{2Y}g(k_T)[\theta(y+Y)-\theta(y-Y)].\]

For minijet gluons: 
\[g(k_T)=N_0 exp\{-\f{k_T}{k_0}\},\]
\[m_T^2\approx \f{3\pi\alpha_s}{R_A^2}\f{N_G}{2Y}.\]
Take $\alpha=0.4$, with minijet rapidity density given by HIJING
$\f{N_G}{2Y}=300$, we get $\mu\sim 4.5 fm^{-1}$;

For minijet $+$ soft gluons:
\[g(k_T)=N_0 (exp\{-\f{k_T}{k_0}\}+\f{40}{3}exp\{-\f{2k_T}{k_0}\}),\]
\[m_T^2\approx \f{92}{169}\f{3\pi\alpha_s}{R_A^2}\f{N_G}{2Y}.\]
For $\alpha=0.4$, with rapidity density $\f{N_G}{2Y}=1300$, we get $\mu\sim 7
fm^{-1}$. 

We'll use these values$^a$ to calculate strangeness production at RHIC energies
with only minijets and minjets+soft gluons and show the important role of soft
gluons in the strangeness production.

\begin{itemize}
\item[a.]
There are uncertainties in the screening because color screening is
intrinsically non-perturbative.
\end{itemize}
\vfill\eject

\section{Strangeness production at RHIC energies}

By following the time evolution of the strangeness and gluon content, we can
monitor the chemical equilibration process. The strangeness to 
gluon ratio is a very important variable for indicating the degree of chemical
equilibration. With chemical equilibrium, we'll have
\[n_s=2\times 3\times 2\times \int \f{d^3p}{(2\pi)^3}\f{1}{e^{\beta E}+1},\]
\[n_g=2\times 8\times \int \f{d^3p}{(2\pi)^3}\f{1}{e^{\beta E}-1},\]
and the ratio
\[n_s/n_g=\f{3}{4}\times \f{3}{4}=\f{9}{16}=0.5625\,.\]
Classically, i.e., without Fermi or Bose statistics, we have $n_s/n_g=3/4=0.75$.

\vspace*{9cm}
\includegraphics{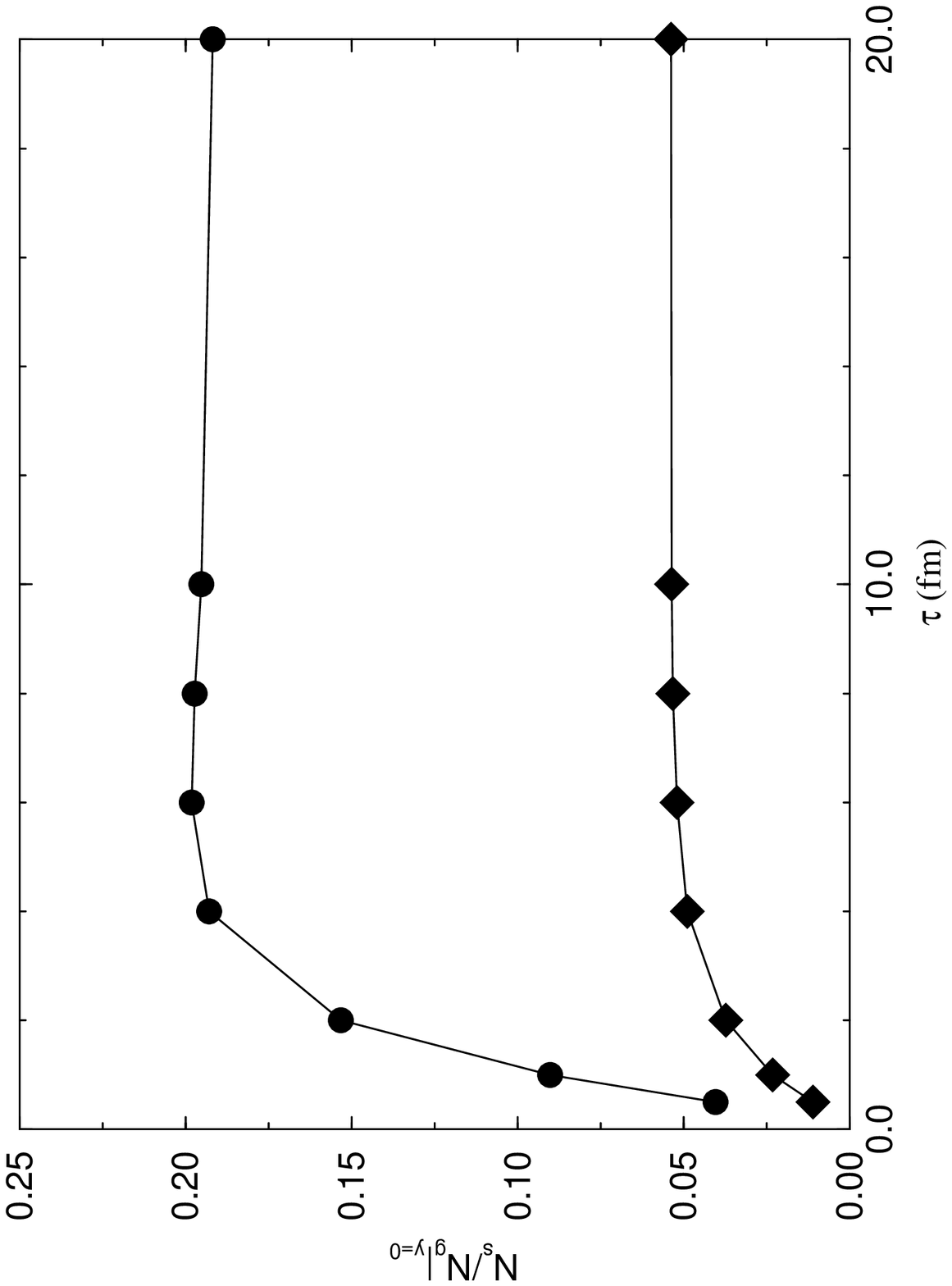}
\vskip -70pt
\begin{minipage}[t]{11cm}
\noindent \bf Fig.2.  \rm
Strangeness to gluon ratio as a function of proper time.
Filled diamonds are for 200 HIJING events with only minijet gluons, screening
mass $\mu=4.5fm^{-1}$, and $\Delta y=1$ around central rapidity; filled circles
are for 30 HIJING events with both minijet gluons and soft gluons,
$\mu=7fm^{-1}$, and $\Delta y=1$. 
\end{minipage}
\vskip 4truemm

Fig. 2 is the time evolution of the strangeness to gluon ratio. We see that
with the soft beam jet gluons, we have a much higher $N_s/N_g$ than that with
only
minijet gluons, because the phase space distribution has changed. Fig. 3 gives
us the Mandelstam s distributions which reflect the change in the phase space
distribution. We see that with the soft gluons, we get an s distribution that
has a lower peak and hence more emphasis on the low energy part of the
strangeness production cross section which is peaked near threshold. So, with
the soft gluons, we have larger strangeness to gluon ratio.

\vspace*{10cm}
\includegraphics{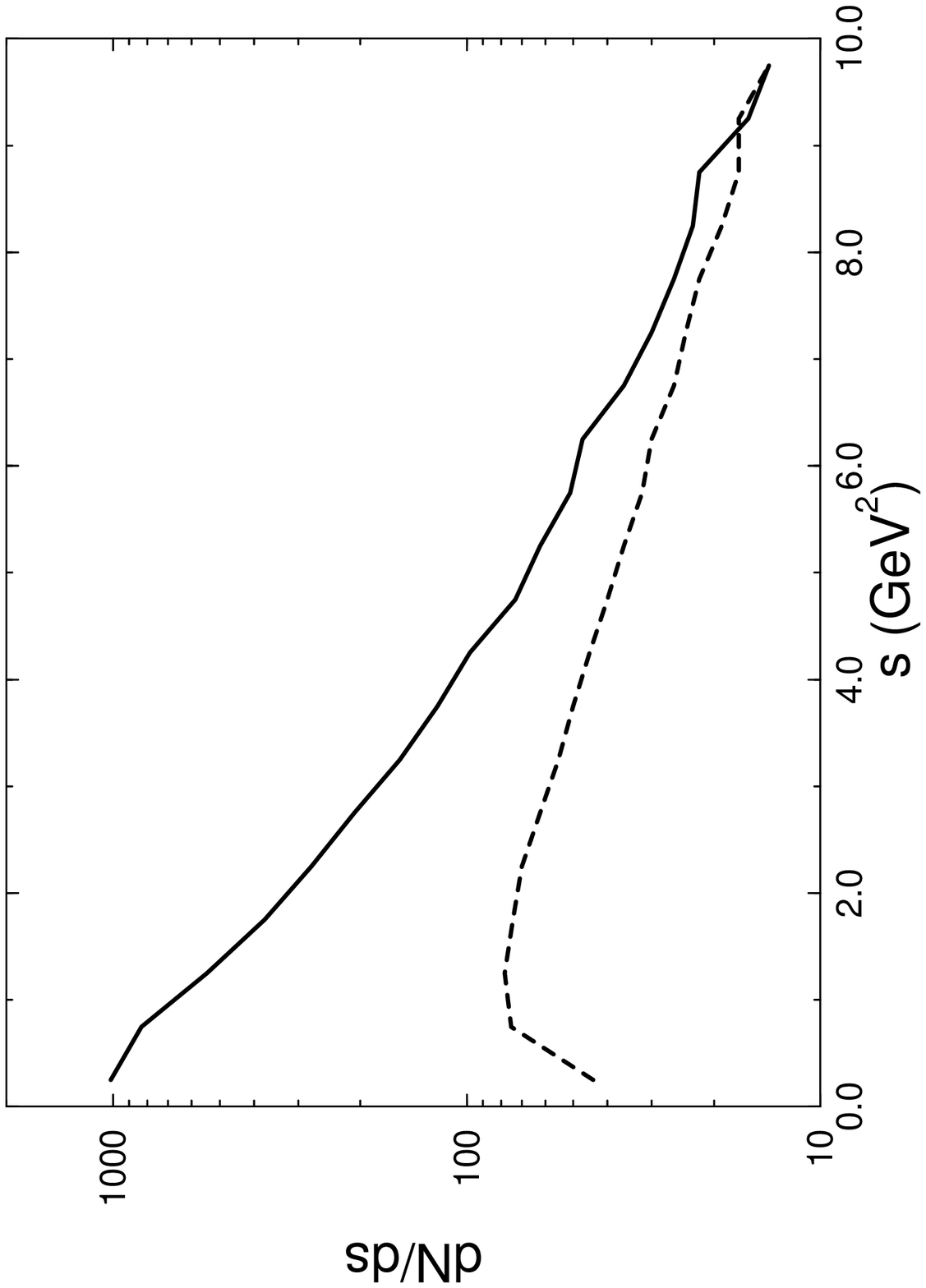}
\vskip -70pt
\begin{minipage}[t]{11cm}
\noindent \bf Fig.3.  \rm
The s distribution of collisions. The dashed curve is $\f{dN}{ds}$ ($N$ is the number
of collisions per event)
 for the case with only minijet gluons ($\mu=4.5 fm^{-1}$). The solid curve is $\f{dN}{ds}$ for
minijet $+$ soft gluons with $\mu=7fm^{-1}$.
\end{minipage}
\vskip 4truemm

For the case with both mini and soft gluons, a more detailed look at the time
evolution of rapidity density of strangeness
and gluons (see Fig. 4 for the result of 30 HIJING events ($\mu=7fm^{-1}$,
$\Delta y=1$)) shows that at central rapidity, they both increase with
proper time. Before 4fms, we get a
rapid increase of the strangeness content, but after 6fms, the gluon content
rises slowly because of the formation of new particles and the scattering of
old particles, but strangeness content freezes out, because the density is too
low for the strangeness production process.

Fig. 5 shows the strangeness production and annihilation rates. 
If chemical equilibrium can be achieved, the annihilation rate
should equal the production rate at some time and then they evolve together
till freezeout. This is not the case here. Before 5fm, the production rate is
much higher than the annihilation rate. After 5fm, the energy momentum tensor
in the local rest frame doesn't have dominant diagonal elements indicating the
system has already frozen
 out. The system can not be
considered thermal at this later stage of evolution.
 
\vspace*{9.5cm}
\includegraphics{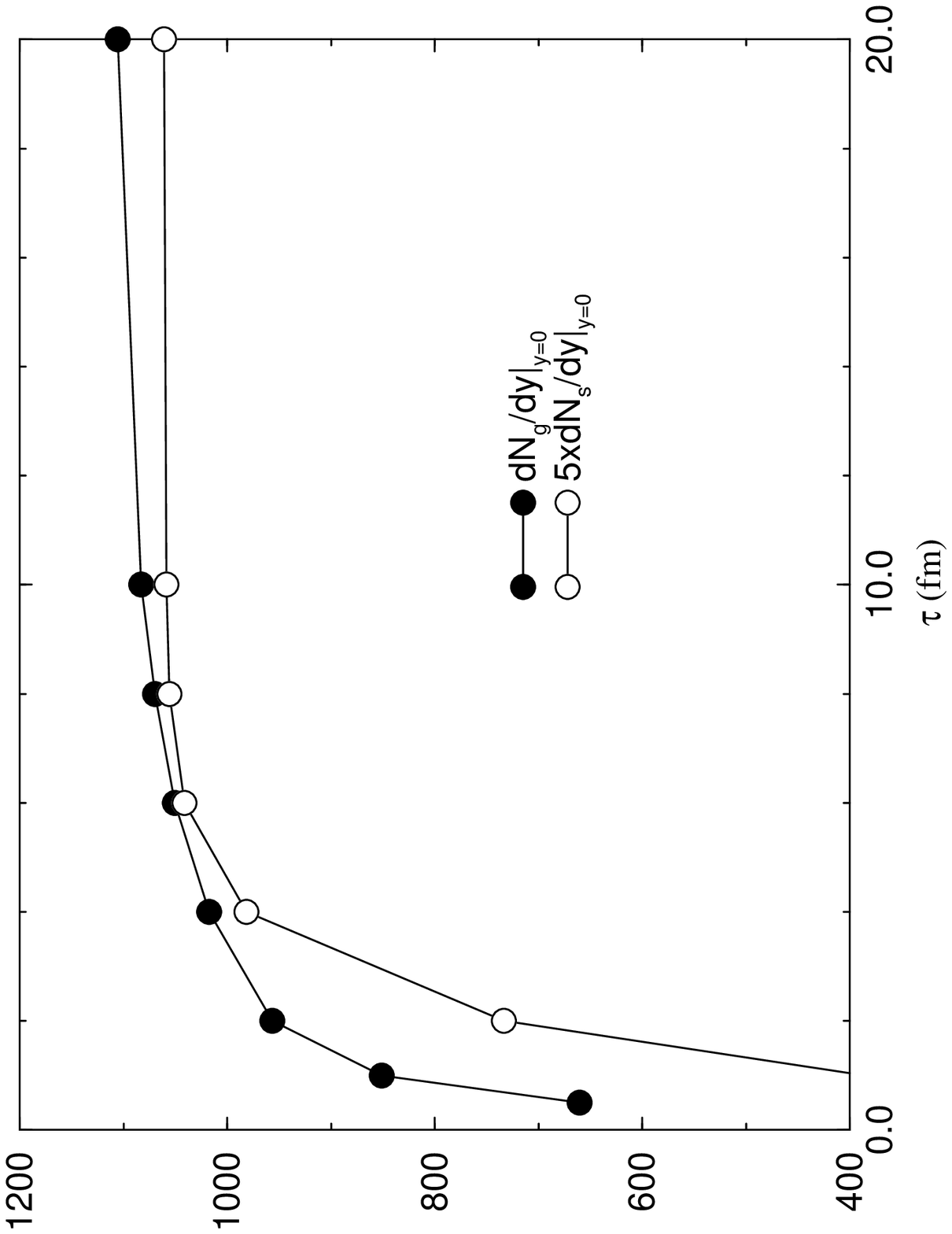}
\vskip -70pt
\begin{minipage}[t]{11cm}
\noindent \bf Fig.4.  \rm
Time developement of rapidity density of gluon and strangeness content per
event.
\end{minipage}
\vskip 4truemm

\vspace*{9.5cm}
\includegraphics{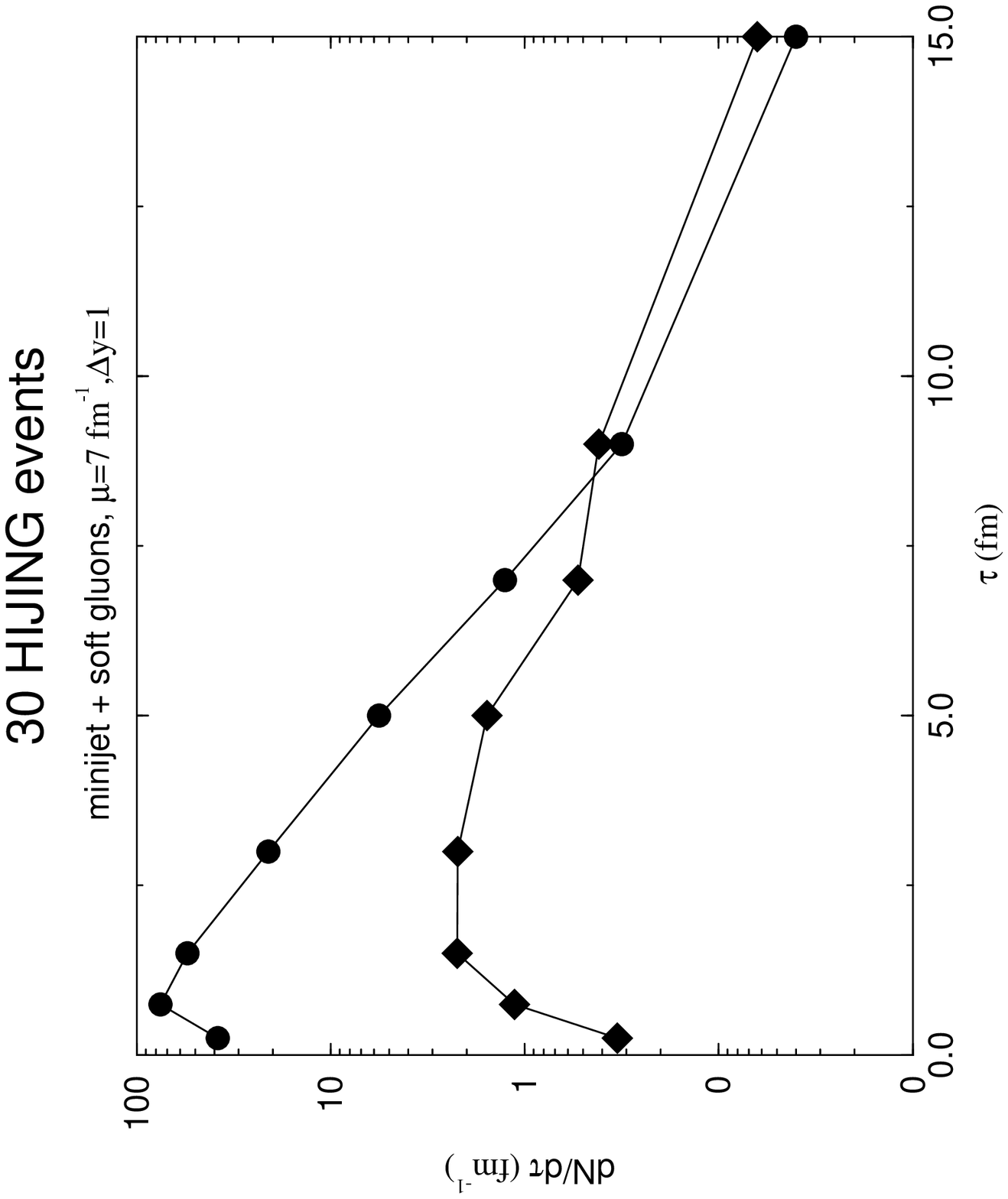}
\vskip -70pt
\begin{minipage}[t]{11cm}
\noindent \bf Fig.5.  \rm
Strangeness production and annihilation rate as a function of time. The
collisions are those with at least one produced particle in the central
rapidity bin. Filled circles are for the
$gg\rightarrow s\bar{s}$ and filled diamonds are for the $s\bar{s}\rightarrow gg$. 
\end{minipage}
\vskip 4truemm

\section{Conclusions}
 
Both the strangeness gluon ratio and strangeness production and annihilation
rates indicate that in Au-Au collisions at RHIC, chemical equilibrium is not
achieved. Soft gluons play an important role in the strangeness production at
RHIC as the minijet gluons and soft gluons together produce a strangeness to
gluon
ratio around 4 times larger than that produced by the minijet gluons alone.

At this stage, we don't have inelastic scatterings, and in addition,
hadronization
may well modify the strangeness to gluon ratio, further investigations are
necessary to address this issue.
 
\vskip 10pt
\noindent \large {\bf Acknowledgement}\normalsize
\vskip 10pt
\noindent
B.Z. likes to thank M. Asakawa and Z. Lin for helpful discussions. 
 This work is supported by the Director, Office of
Energy Research, Division of Nuclear Physics of the Office of High Energy and
Nuclear Physics of the U.S. Department of Energy under Contract
No. DE-FG02-93ER40764 and U.S. Department of Energy Contract DE-FG-02-92
ER40699.     


\vfill\eject 

\begin{thebibliography}{99}\parindent=8truemm
\itemsep -1mm

\bibitem{bib1} J. Rafelski \& B. M\"{u}ller, {\it PRL} {\bf 48}
(1982) 1066, {\it PRL} {\bf 56} (1986) 2334; T.Biro \& J. Zimanyi, {\it PL}
{\bf
113B} (1982) 6, {\it Nucl. Phys.} {\bf A395} (1983) 525; T. Matui, B. Svetitsky
\& L. McLerran, {\it PRD} {\bf 34} (1986) 783, {\it PRD} {\bf 34} (1986) 2047;
P. Koch, B. M\"{u}ller \& J. Rafelski, {\it PR} {\bf 142} (1986)
147.

\bibitem{bib2} X-N. Wang \& M. Gyulassy, {\it PRD} {\bf 44} (1991) 3501.

\bibitem{bib3} P. L\'{e}vai \& X-N. Wang, {\it hep-ph/9504214}.

\bibitem{bib4} K. Geiger, {\it PRD} {\bf 46} (1992) 4965, {\it PRD} {\bf 46}
(1992) 4986, {\it Proceedings of the Workshop on Pre-equilibrium Parton
Dynamics}, (1993 edited by X-N. Wang), 61.

\bibitem{bib5} T. Biro, B. M\"{u}ller \& X-N. Wang (\it PLB) {\bf
283} (1992) 171.

\end{thebibliography}
\end{document}